\documentclass[astrosymb, twocolumn]{aastex631}

\usepackage{booktabs}
\usepackage{dsnr}

\shorttitle{MIRI Spectra of a Quasar Wind}
\shortauthors{Rupke et al.}

\newcommand{\ftm}{F2M1106}
\newcommand{\jwst}{\emph{JWST}}

\graphicspath{{./}{figures/}}

\begin{document}

\title{First Results from the JWST Early Release Science Program Q3D:
  Benchmark Comparison of Optical and Mid-infrared Tracers of a Dusty, Ionized Red Quasar Wind at $z$=0.435}

\correspondingauthor{David Rupke}
\email{drupke@gmail.com}

\author[0000-0002-1608-7564]{David S. N. Rupke}
\affiliation{Department of Physics, Rhodes College, 2000 N. Parkway, Memphis, TN 38112, USA}

\author[0000-0003-2212-6045]{Dominika Wylezalek}
\affiliation{Zentrum für Astronomie der Universität Heidelberg, Astronomisches Rechen-Institut, Mönchhofstr 12-14, D-69120 Heidelberg, Germany}

\author[0000-0001-6100-6869]{Nadia L. Zakamska}
\affiliation{Department of Physics and Astronomy, Bloomberg Center, Johns Hopkins University, 3400 N. Charles St., Baltimore, MD 21218, USA}
\affiliation{Institute for Advanced Study, Princeton, NJ 08540, USA}

\author[0000-0002-3158-6820]{Sylvain Veilleux}
\affiliation{Department of Astronomy and Joint Space-Science Institute, University of Maryland, College Park, MD 20742, USA}

\author[0000-0002-6948-1485]{Caroline Bertemes}
\affiliation{Zentrum für Astronomie der Universität Heidelberg, Astronomisches Rechen-Institut, Mönchhofstr 12-14, D-69120 Heidelberg, Germany}

\author[0000-0001-7572-5231]{Yuzo Ishikawa}
\affiliation{Department of Physics and Astronomy, Bloomberg Center, Johns Hopkins University, 3400 N. Charles St., Baltimore, MD 21218, USA}

\author[0000-0003-3762-7344]{Weizhe Liu}
\affiliation{Steward Observatory, University of Arizona, 933 North Cherry Avenue, Tucson, AZ 85721, USA}

\author[0000-0002-4419-8325]{Swetha Sankar}
\affiliation{Department of Physics and Astronomy, Bloomberg Center, Johns Hopkins University, 3400 N. Charles St., Baltimore, MD 21218, USA}

\author[0000-0002-0710-3729]{Andrey Vayner}
\affiliation{Department of Physics and Astronomy, Bloomberg Center, Johns Hopkins University, 3400 N. Charles St., Baltimore, MD 21218, USA}

\author{Hui Xian Grace Lim}
\affiliation{Department of Physics, Rhodes College, 2000 N. Parkway, Memphis, TN 38112, USA}

\author{Ryan McCrory}
\affiliation{Department of Physics, Rhodes College, 2000 N. Parkway, Memphis, TN 38112, USA}

\author[0009-0007-7266-8914]{Grey Murphree}
\affiliation{Department of Physics, Rhodes College, 2000 N. Parkway, Memphis, TN 38112, USA}
\affiliation{Institute for Astronomy, University of Hawai'i, Honolulu, HI, 96822, USA}

\author{Lillian Whitesell}
\affiliation{Department of Physics, Rhodes College, 2000 N. Parkway, Memphis, TN 38112, USA}

\author[0000-0001-9495-7759]{Lu Shen}
\affiliation{CAS Key Laboratory for Research in Galaxies and Cosmology, Department of Astronomy, University of Science and Technology of China, Hefei, Anhui 230026, China}
\affiliation{School of Astronomy and Space Science, University of Science and Technology of China, Hefei 230026, China}
\affiliation{Department of Physics and Astronomy, Texas A\&M University, College Station, TX 77843-4242 USA}

\author[0000-0003-4286-5187]{Guilin Liu}
\affiliation{CAS Key Laboratory for Research in Galaxies and Cosmology, Department of Astronomy, University of Science and Technology of China, Hefei, Anhui 230026, China}
\affiliation{School of Astronomy and Space Science, University of Science and Technology of China, Hefei 230026, China}

\author[0000-0003-2405-7258]{Jorge K. Barrera-Ballesteros}
\affiliation{Instituto de Astronomía, Universidad Nacional Autónoma de México, AP 70-264, CDMX 04510, Mexico}

\author[0000-0001-8813-4182]{Hsiao-Wen Chen}
\affiliation{Department of Astronomy \& Astrophysics, The University of Chicago, 5640 South Ellis Avenue, Chicago, IL 60637, USA}

\author[0009-0003-5128-2159]{Nadiia Diachenko}
\affiliation{Department of Physics and Astronomy, Bloomberg Center, Johns Hopkins University, 3400 N. Charles St., Baltimore, MD 21218, USA}

\author[0000-0003-4700-663X]{Andy D. Goulding}
\affiliation{Department of Astrophysical Sciences, Princeton University, 4 Ivy Lane, Princeton, NJ 08544, USA}

\author[0000-0002-5612-3427]{Jenny E. Greene}
\affiliation{Department of Astrophysical Sciences, Princeton University, 4 Ivy Lane, Princeton, NJ 08544, USA}

\author[0000-0003-4565-8239]{Kevin N. Hainline}
\affiliation{Steward Observatory, University of Arizona, 933 North Cherry Avenue, Tucson, AZ 85721, USA}

\author{Fred Hamann}
\affiliation{Department of Physics \& Astronomy, University of California, Riverside, CA 92521, USA}

\author[0000-0001-8813-4182]{Timothy Heckman}
\affiliation{Department of Physics and Astronomy, Bloomberg Center, Johns Hopkins University, 3400 N. Charles St., Baltimore, MD 21218, USA}

\author[0000-0001-9487-8583]{Sean D. Johnson}
\affiliation{Department of Astronomy, University of Michigan, Ann Arbor, MI 48109, USA}

\author[0000-0003-0291-9582]{Dieter Lutz}
\affiliation{Max-Planck-Institut für Extraterrestrische Physik, Giessenbachstrasse 1, D-85748 Garching, Germany}

\author[0000-0001-6126-5238]{Nora Lützgendorf}
\affiliation{European Space Agency, Space Telescope Science Institute, Baltimore, MD, USA}

\author[0000-0002-1047-9583]{Vincenzo Mainieri}
\affiliation{European Southern Observatory, Karl-Schwarzschild-Straße 2, D-85748 Garching bei München, Germany}


\author[0000-0001-5783-6544]{Nicole P. H. Nesvadba}
\affiliation{Université de la Côte d'Azur, Observatoire de la Côte d'Azur, CNRS, Laboratoire Lagrange, Bd de l'Observatoire, CS 34229, F-06304 Nice cedex 4, France}

\author[0000-0002-3471-981X]{Patrick Ogle}
\affiliation{Space Telescope Science Institute, 3700 San Martin Drive, Baltimore, MD 21218, USA}

\author[0000-0002-0018-3666]{Eckhard Sturm}
\affiliation{Max-Planck-Institut für Extraterrestrische Physik, Giessenbachstrasse 1, D-85748 Garching, Germany}

\begin{abstract}
The \othl\ emission line is the most common tracer of warm, ionized outflows in active galactic nuclei across cosmic time. \jwst\ newly allows us to use mid-IR spectral features at both high spatial and spectral resolution to probe these same winds. Here we present a comparison of ground-based, seeing-limited \oth\ and space-based, diffraction-limited \sufl\ maps of the powerful, kiloparsec-scale outflow in the Type 1 red quasar SDSS J110648.32+480712.3. The \jwst\ data are from the Mid-InfraRed Instrument. There is a close match in resolution between the datasets ($\sim$0\farcs6), in ionization potential of the O$^{+2}$ and S$^{+3}$ ions (35~eV), and in line sensitivity ($1-2\times10^{-17}$ erg~s$^{-1}$~cm$^{-2}$~arcsec$^{-2}$). The \oth\ and \suf\ line shapes match in velocity and line width over much of the 20~kpc outflowing nebula, and \suf\ is the brightest line in the rest-frame 3.5--19.5~$\mu$m range, demonstrating its usefulness as a mid-IR probe of quasar outflows. \oth\ is nevertheless intrinsically brighter and provides better contrast with the point-source continuum, which is strong in the mid-IR. There is a strong anticorrelation of \oth/\suf\ with average velocity, which is consistent with a scenario of differential obscuration between the approaching (blueshifted) and receding (redshifted) sides of the flow. The dust in the wind may also obscure the central quasar, consistent with models that attribute red quasar extinction to dusty winds.
\end{abstract}

\section{Introduction} \label{sec:intro}

Galaxy-scale outflows driven by luminous active galactic nuclei (AGN), and especially by their most powerful representatives -- quasars, are a cornerstone of modern galaxy evolution models \citep{silk98, king03, fabian12}. They have been shown to be ubiquitous in the nearby universe, primarily through observations of the bright \othl\ line that is prominent in the rest-frame optical (e.g., \citealt{2013MNRAS.436.2576L}). This line is strong in AGN-photoionized regions because far-UV photons from the power-law radiation produce significant amounts of doubly ionized oxygen \citep{1987ApJS...63..295V}, which has an ionization energy of 35.12~eV. 

``Q3D: Imaging Spectroscopy of Quasar Hosts with JWST" is an Early Release Science Program on \jwst\ \citep{2022ApJ...940L...7W} to establish infrared diagnostics of quasar-driven outflows and to develop techniques for probing their impact on the quasar host evolution. One of the program's major goals is to extend the detailed observations of \oth\ to redshifts of up to $z \sim 3$ \citep{2022ApJ...940L...7W, veilleux23, vayner23a}, in part to benchmark the effects of these outflows at Cosmic Noon -- the epoch of peak star formation and quasar activity \citep{boyle98}. Another key goal of this program is to probe these outflows at high spatial and spectral resolution using near-IR and mid-IR capabilities of \jwst. The revolutionary combination of high resolution and sensitivity at rest-frame mid-IR wavelengths opens new windows to observe and characterize quasar outflows and their impact on the quasar host galaxies.

The nearest quasar in the Q3D sample is a Type 1 quasar with $z = 0.4350$ and bolometric luminosity $L_\mathrm{bol}=10^{46}-10^{47}$ erg~s$^{-1}$  \citep{shen23}. SDSS J110648.32$+$480712.3 was selected as a red quasar from a cross-correlation of the Faint Images of the Radio Sky at Twenty Centimeters (FIRST; \citealt{1995ApJ...450..559B}) and Two Micron All Sky Survey (2MASS; \citealt{skrutskie06}) surveys \citep{2012ApJ...757...51G}. \citet{2012ApJ...757...51G} give it the catalog label F2M110648.35$+$480712.3---\ftm\ for short---and report $J - K = 2.31$ mag, $R - K = 4.01$ mag, and $\ebv = 0.44$ mag, which are at the blue edge (in $R - K$) of the red quasar color-color selection region. Sloan Digital Sky Survey (SDSS) and Wide-field Infrared Survey Explorer (WISE) photometry of \ftm\ give an optical-to-mid-IR color of $r_\mathrm{AB} - W4_\mathrm{Vega}=12.7$ mag. Following \citet{2015MNRAS.453.3932R}, $A_V = R_V\times\ebv = 2.87*0.44 = 1.19$ mag. Thus, \ftm\ is $\Delta A_V = 1$ mag below the threshold for an extremely red quasar (ERQ; \citealt{2015MNRAS.453.3932R}).

\ftm\ entered the Q3D sample thanks to the discovery of a fast, powerful quasar outflow clearly detected in \oth\ and extending out to distances over 10~kpc from the nucleus \citep{shen23}. Due to its proximity, we can use \ftm\ to explore tracers of this quasar outflow in the rest-frame near- and mid-IR.

A prominent line that appears in the rest-frame, mid-IR \jwst\ spectrum of \ftm\ is \sufl. The energies required to produce S$^{+3}$ and O$^{+2}$ are almost identical (34.79~eV and 35.12~eV, respectively). Their critical densities are larger than typical densities in the narrow-line region: $6.8\times10^5$~cm$^{-3}$ for \othl\ \citep{2006agna.book.....O} and $5.4\times10^4$~cm$^{-3}$ for \sufl\ \citep{2002ApJ...566..880G}. \suf\ and \oth\ should thus probe the same gas phase, assuming that they arise from the same region that is illuminated by the intense UV radiation of the quasar. Both lines can be excited by starlight, AGN photoionization, and shocks \citep{2003A&A...403..829V, 2004ApJS..153....9G, 2008ApJS..178...20A}. Comparisons to other lines can distinguish among these possibilities, though quasar or shock photoionization are most likely in the \ftm\ outflow (as we discuss in Section~\ref{sec:discuss}).

One key difference is that the lines experience significantly different extinction ($A_\lambda/A_V = 1.10$ vs. 0.14 for \oth\ and \suf, respectively). Thus, \oth/\suf\ will decrease by a factor of $e$ under an $A_V$ of unity, and the ratio is sensitive to extinction. \sufl\ falls within the reach of the broad 9.7~$\mu$m silicate absorption feature, which is incorporated in this estimate.

Fortuitously, the delivered \jwst-MIRI/MRS resolution at 15.1~$\mu$m, the observed wavelength of this line, is 0\farcs6--0\farcs7, comparable to the resolution of the Gemini Multi-Object Spectrograph (GMOS) \oth\ data cube (0\farcs6). We will show that differential beam-smearing and point spread function (PSF) mismatch are not a concern. Comparing these two lines can thus serve to simultaneously benchmark (1) the performance of \jwst\ at these wavelengths for this science and (2) the use of \sufl\ as a rest-frame mid-IR tracer of quasar outflows that is potentially identical to \othl\ in the rest-frame optical.

We present the \jwst\ observations, reduction, and analysis, and recap the ground-based \oth\ data, in Section~\ref{sec:obs}. In Section~\ref{sec:results} we present the maps of \suf\ line properties and compare to those of \oth. We discuss the implications of this comparison in Section~\ref{sec:discuss}. We conclude in Section~\ref{sec:conclude} and briefly assess the PSFs the Appendix. In all calculations, we assume a flat $\Lambda$ cosmology with $\Omega_m=0.3$ and $H_0=70$~\kms~Mpc$^{-1}$. $v_{50\%}$ and $W_{80\%}$ denote the velocity at which 50\%\ of line flux accumulates and the velocity width containing 80\%\ of line flux, respectively \citep{2013MNRAS.436.2576L,2014MNRAS.442..784Z}. The redshift of the quasar (0.4350) is determined using stellar position-velocity diagrams from ground-based optical integral-field spectroscopy (D. S. N. Rupke et al., in preparation).

\begin{figure*}
    \centering
    \includegraphics[width=1.15\textwidth]{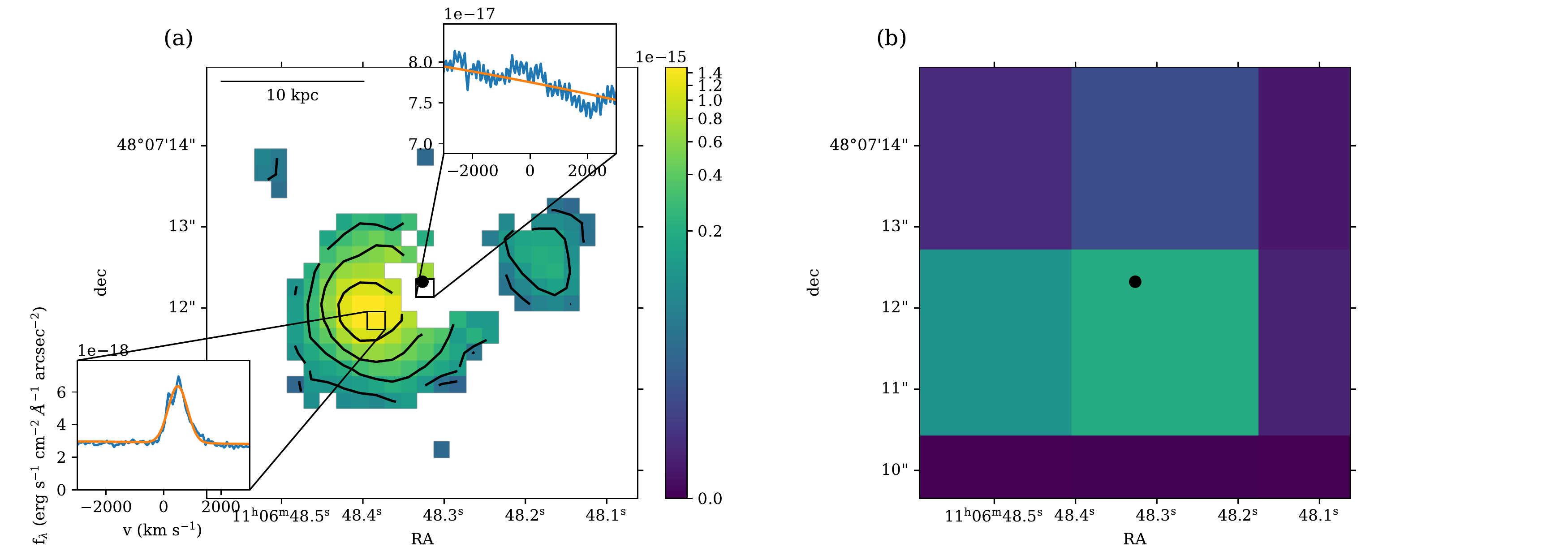}
    \caption{(a) Total \sufl\ model surface brightness in each 0\farcs2 spaxel, in erg~s$^{-1}$~cm$^{-2}$~arcsec$^{-2}$. A black, filled circle locates the quasar. The inset shows the spectrum of, and fit to, two spaxels. Surface brightness contours begin at 1$\times$10$^{-15}$ erg~s$^{-1}$~cm$^{-2}$~arcsec$^{-2}$ and descend by factors of 2. (b) Model flux reprojected to 2\farcs3 spaxels, to illustrate the 10$\times$ improved resolution of MIRI/MRS compared to the sampling of the Short-High module of the InfraRed Spectrograph on board {\em Spitzer}.}
    \label{fig:jwst-vs-spitzer}
\end{figure*}

\section{Observations, Data Reduction, and Fitting} \label{sec:obs}

\subsection{\jwst}

We observed \ftm\ with the Medium-Resolution Spectrometer (MRS) mode of the Mid-InfraRed Instrument (MIRI; \citealt{2015PASP..127..584R}) on \jwst\ as part of the Q3D program (DD-ERS-1335, PI: Wylezalek). We acquired data at each of the three grating settings (Short, Medium, and Long bands) through Channels 1--4, yielding a data cube with a continuous wavelength coverage from 4.9 to 27.9~$\mu$m. We acquired dedicated background exposures, and dithered in a four-point pattern to increase the sampling and spatial resolution of the drizzled, postpipeline data cube.

We reduce the data using version 1.9.5 of the {\jwst} pipeline \citep{bushouse_howard_2023_7692609}. In \texttt{spec2}, we apply both initial and residual fringe corrections. In \texttt{spec3}, we perform image background subtraction and outlier detection and use the \texttt{emsm} algorithm for weighting during cubebuilding. We build the cube discussed here for a single band; the cube thus has 0\farcs2$\times$0\farcs2 spaxels. We register the data to the SDSS coordinate using the centroid from a 2D Gaussian fit to the wavelength-collapsed cube.

Here we present spatially resolved data from the Medium band of Channel 3 (i.e., Channel 3B), which contains the redshifted \sufl\ emission line. Channel 3 has 0\farcs245 pixels and a slice width of 0\farcs387. The reduced cube has a spatial resolution of 0\farcs6--0\farcs7. A Gaussian fit to the 1D radial profile of the collapsed Channel 3B cube has a best-fit FWHM of 0\farcs65. A small amount of power ($\sim$10\%) is in non-Gaussian wings beyond $r\sim3$ pixels (see the Appendix). The cube is subject to low-frequency ripples in spaxels at and near the quasar center due to the undersampling of the MIRI PSF at these wavelengths. It is also subject to residual high-frequency fringing in the same region.

We also present preliminary results from a 1D spectrum extracted from the Long band of Channel 2 (2C); Channel 3B; and the Short and Medium bands of Channel 4 (4A and 4B). These contain the [\ion{Ne}{6}] 7.65~$\mu$m; [\ion{Ne}{2}] 12.81~$\mu$m and [\ion{Ne}{5}] 14.32~$\mu$m; and [\ion{Ne}{3}] 15.56~$\mu$m features, respectively. We extract this spectrum in a 0\farcs6 circular aperture centered on the peak \suf\ flux, which lies at coordinates 11:06:48.4, $+$48:07:12.0. Presentation of the full MIRI spectrum of this source, spanning all bands of each channel, is pending a full characterization of the data.

We fit the data using \texttt{q3dfit} v1.1.0, a software package designed for integral-field spectra from \jwst. Unlike \oth, \suf\ has a relatively low equivalent width in AGN spectra, including obscured and red quasars, because of the strong thermal emission continuum produced by the AGN-heated dust at these wavelengths. \suf\ is undetected in individual spaxels in the nucleus (Figure~\ref{fig:jwst-vs-spitzer}a) and its equivalent width in a spatially integrated nuclear spectrum is extremely low (Section~\ref{sec:results}). Thus, we do not remove a PSF component from the extended line emission. We use a simple order-3 polynomial and 1--2 Gaussian emission lines to fit the continuum and \suf\ line in each spaxel of Channel 3B. The continuum is dominated by the bright nuclear point source. We convolve each line model with the spectral response before fitting; Channel 3B has a resolving power $R\sim2500$. Fits with peak \suf\ flux below 3$\sigma$ are rejected. To account for undersampling ripples, we compute $\sigma$ both from the line fit covariance matrix and from the rms of the continuum fit in a 200-pixel window around the line and apply the greater of these. The limiting surface brightness of the line fits is approximately 10$^{-17}$ erg~s$^{-1}$~cm$^{-2}$~arcsec$^{-2}$.

We separately fit the individual channels of the 1D spectrum using low-order polynomials (orders 1--3) and 1--2 components per line. We also fit the [\ion{Cl}{2}] 14.37~$\mu$m line to deblend it from [\ion{Ne}{5}].

\subsection{Gemini}

We observed \ftm\ for a total exposure time of 3240~s with the Gemini Multi-Object Spectrograph (GMOS) on Gemini North (GN-2014A-Q-19, PI: Liu). GMOS has 0\farcs2 hexagonal spaxels, almost identical to the sampling of MIRI at these wavelengths. The details of the data reduction and analysis are described in \citet{shen23}. The seeing was 0\farcs4; however, the delivered image quality is coarser. Summing the cube over observed frame 7300--7400~\AA, we fit the 1D radial profile with a Gaussian of FWHM $=0\farcs62$. As was the case for MIRI, there are non-Gaussian wings that carry about 15\%\ power. The MIRI and GMOS PSFs are thus nearly identical (see the Appendix).

We coregister the GMOS data to the MIRI world coordinate system (WCS) using the centroid from a 2D Gaussian fit to the same 7300--7400~\AA\ sum.

The maps presented below have 0\farcs05 sampling and a 5$\sigma$ cut applied to the line flux. The limiting surface brightness is 2$\times$10$^{-17}$ erg~s$^{-1}$~cm$^{-2}$~arcsec$^{-2}$. However, when directly comparing to the MIRI data, we {\tt reproject} \citep{thomas_robitaille_2023_7950746} the GMOS data onto the same WCS.

\section{Results} \label{sec:results}

In Figure \ref{fig:jwst-vs-spitzer}a, we show the fit results as a surface brightness map of \suf. The emission peaks 0\farcs8 SE and 1\farcs5 NW (4.7~kpc SE and 8.3~kpc NW) of the  nucleus. The nucleus is determined from our centroid fits to the MIRI and GMOS cubes, collapsed over wavelength (Section~\ref{sec:obs}). The insets show the two-component profile near the peak line flux, as well as the strong continuum flux and low- and high-frequency ripples that prevent the detection of low-intensity \suf\ in and near the nucleus. In Figure \ref{fig:jwst-vs-spitzer}b, we demonstrate how this same map would appear if observed by {\it Spitzer} with the SH module at 10$\times$ lower spatial resolution (and 4$\times$ lower spectral resolution; \citealt{2021A&A...656A..57L}; \citealt{https://doi.org/10.26131/irsa487}). This starkly illustrates the power of {\it JWST} spatial resolution in the mid-IR to resolve quasar outflows.

\begin{figure*}
  \centering
  \includegraphics[width=\textwidth]{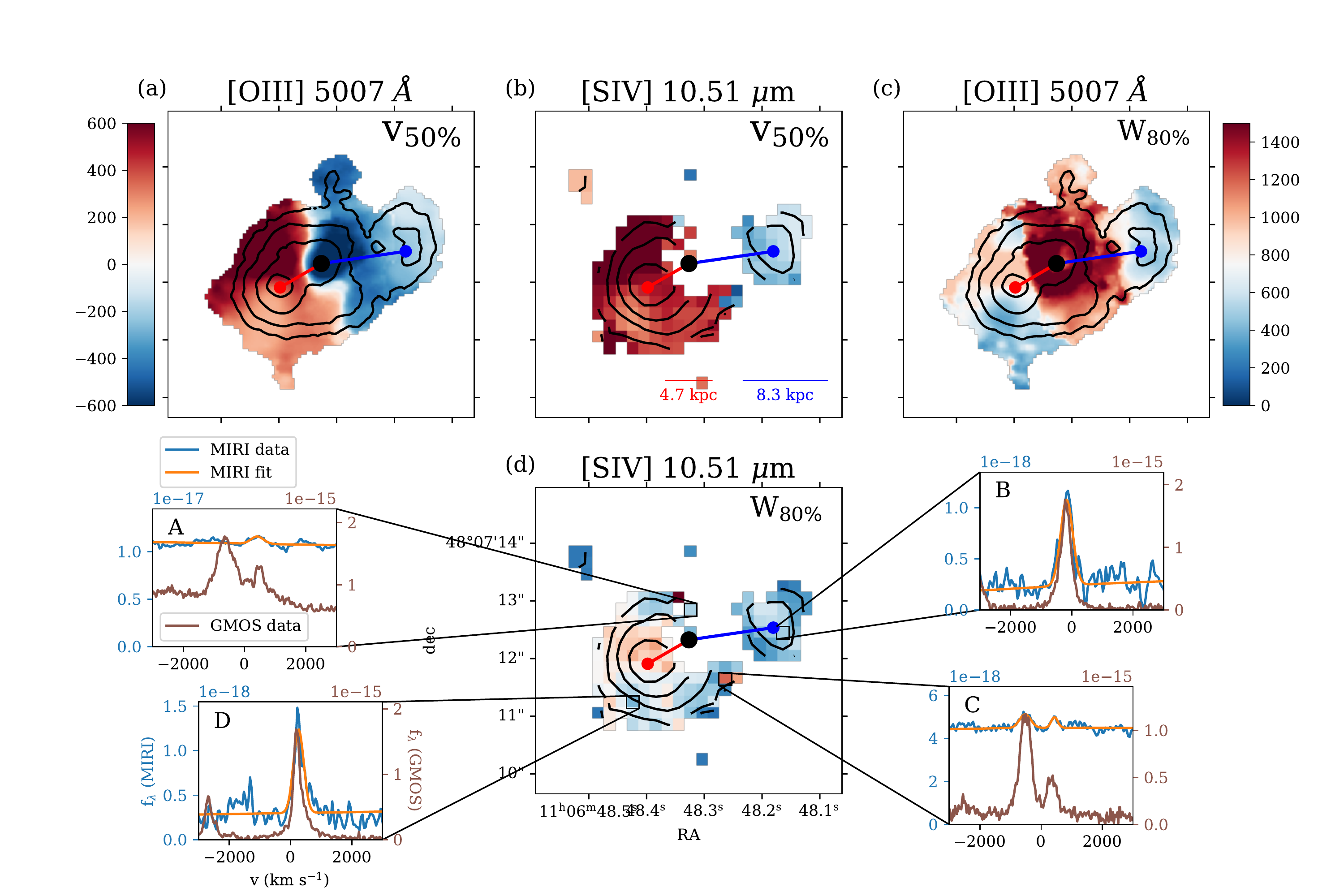}
    \caption{(a) \oth\ $v_\mathrm{med}$ from Gemini-GMOS observations \citep{shen23}. Surface brightness contours begin at 5$\times$10$^{-15}$ erg~s$^{-1}$~cm$^{-2}$~arcsec$^{-2}$ and descend by factors of 2. The quasar is indicated with a black circle. (b) \suf\ $v_{50\%}$, in \kms, with surface brightness contours as in Figure~\ref{fig:jwst-vs-spitzer}. The \suf\ surface brightness peaks are plotted as red and blue circles, and they are connected to the quasar by red and blue lines. The lengths of these lines are indicated below. The velocity scale is identical to panel (a). (c) \oth\ $W_{80\%}$. Note the N-to-SW band of high $W_{80\%}$ that is not present in \suf. (d) \suf\ $W_{80\%}$. The velocity scale is identical to panel (c). Insets show MIRI and GMOS spectra in several representative spaxels with fits to the MIRI data overlaid. Flux densities are in erg~s$^{-1}$~cm$^{-2}$~arcsec$^{-2}$. The left-side y-axis shows MIRI $f_\lambda$; the right-side y-axis is GMOS $f_\lambda$.}
    \label{fig:s4-on-o3}
\end{figure*}

\begin{figure*}
  \centering
  \includegraphics[width=\textwidth]{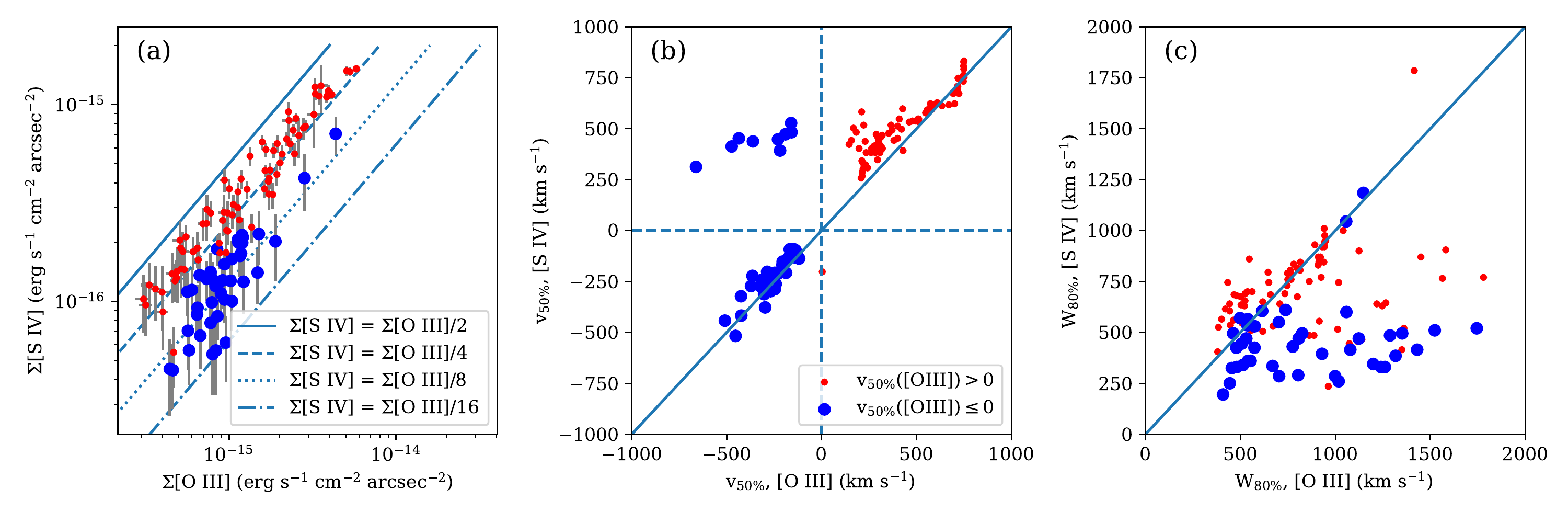}
    \caption{(a) \suf\ vs. \oth\ surface brightnesses. Spaxels with \oth\ $v_{50\%}>0$~\kms\ are in red; blueshifted spaxels are in blue. Lines of constant flux ratios are overplotted. Blueshifted spaxels have lower \oth/\suf. Gray lines represent 1$\sigma$ error bars. (b) \suf\ vs. \oth\ $v_{50\%}$. The diagonal solid line is a line of equality; the dashed lines locate $v=0$~\kms. Several points show blueshifted \oth\ compared to \suf; these are in the N-to-SW band of the nebula where the red and blue sides of the outflow overlap and only one component of \suf\ is detected. Uncertainties are of order 10s of \kms. (c) \suf\ vs. \oth\ $W_{80\%}$. The solid line shows equality. Again, spaxels with larger \oth\ $W_{80\%}$ compared to \suf\ indicate locations where more \oth\ components are detected, and also where a faint, broad base of the line is undetected in \suf. Uncertainties are of order 100 \kms.}
    \label{fig:s4-vs-o3}
\end{figure*}

We make a direct comparison to the ground-based \othl\ observations in Figures~\ref{fig:s4-on-o3}--\ref{fig:s4-vs-o3}. The correspondence is excellent in most regions of the nebula. The \suf\ and \oth\ nebulae peak in the same locations on either side of the quasar, and the morphology is very similar where both are detected. The red- and blue-shifted sides of the bipolar outflow extend to the SE and NW, respectively, in Figure \ref{fig:s4-on-o3}a--b. The velocities and line widths of the two lines are identical in many spaxels. 

Several differences between the tracers reflect observational and intrinsic effects. The first can be seen in the central regions. A band of high \oth\ $W_{80\%}$ runs from the N to the SW (Figure~\ref{fig:s4-on-o3}c). This band arises due to the overlap of the receding and approaching sides of the outflow. In these spaxels, both red- and blue-shifted components are detected \citep{shen23}. In contrast, in many of these spaxels \suf\ remains undetected, or only one component is detected. An example of undetected \suf\ is the central spaxel (Figure~\ref{fig:jwst-vs-spitzer}a); an example of only one component being detected is spaxel A of Figure~\ref{fig:s4-on-o3}d. The \suf\ line widths in this region are correspondingly smaller. In a small number of spaxels both sides of the outflow are detected in \suf\ (e.g., spaxel C of Figure~\ref{fig:s4-on-o3}d); in these cases the line widths are indeed more comparable to those of \oth. The differences between the observed \suf\ and \oth\ velocity distributions in this central, N-to-SW band are reflected in line widths $W_{80\%}$ that are larger  in \oth\ compared to \suf\ by up to 1000~\kms\ and $v_{50\%}$ values that are more blueshifted in \oth\ than in \suf\ by up to 1000~\kms\ (Figure~\ref{fig:s4-vs-o3}b--c).

We attribute the missing \suf\ at some velocities in these central spaxels to the relative weakness of \suf\ compared to \oth, both in absolute terms and in comparison to the nuclear continuum. Both observations reach similar surface brightness sensitivities (Section~\ref{sec:obs}). However, where both lines are detected, \suf\ is 2--16$\times$ weaker than \oth\ (Figure~\ref{fig:s4-vs-o3}a). The mid-IR continuum is dominated by the nuclear point source, which in turn affects the central $\sim$1\arcsec\ because of the PSF. In a 0\farcs6 diameter nuclear spectrum, we measure rest-frame equivalent widths $W_\mathrm{eq} = 16.8\pm0.4$~\AA\ and $9\pm2$~\AA\ for \oth\ and \suf, respectively. A direct comparison is possible in velocity space; we estimate $700\pm10$~\kms\ and $18\pm3$~\kms, respectively. Thus, the nuclear equivalent width ratio of \oth\ to \suf\ is 39. Examples of how this difference in the nuclear $W_\mathrm{eq}$ affects the detectability of \suf\ vs. \oth\ are spaxels A and C in Figure~\ref{fig:s4-on-o3}d. Farther from the central point source, the PSF contamination of the underlying continuum is much smaller (spaxels B and D in Figure~\ref{fig:s4-on-o3}d).

A further effect that may contribute to the difference in $W_{80\%}$ is insensitivity to a faint but broad line base that is undetected in \suf\ \citep{2014MNRAS.442..784Z}. For instance, spaxels B and D (Figure~\ref{fig:s4-on-o3}d) show very faint line wings in \oth\ that remain undetected in \suf. These spaxels are near the signal-to-noise ratio (S/N) $\sim$ 10 threshold identified by \citet{2014MNRAS.442..784Z} for robustly detecting a broad base. In this case $W_{80\%}$ is largely unaffected, but in the central band the S/N of \suf\ is certainly affecting $W_{80\%}$.

The second key difference between the two tracers is seen in the range of observed line flux ratios, $\Sigma(\oth)/\Sigma(\suf)$. In Figure~\ref{fig:s4-vs-o3}a, we show that the fluxes from the two lines are generally correlated with each other, but that there is a range of ratios. The range is 2--16, but the ratio appears to depend on velocity. In Figure~\ref{fig:s4o3-vs}a, we show more clearly that there is a strong, significant correlation of log($\Sigma(\oth)/\Sigma(\suf)$) with $v_{50\%}(\oth)$ ($r = -0.75$). It also correlates significantly with the differences between the \oth\ and \suf\ line widths, though more weakly ($r = 0.50$; Figure~\ref{fig:s4o3-vs}b). It does not correlate significantly with projected galactocentric radius $R$ (Figure~\ref{fig:s4o3-vs}c).

\section{Discussion} \label{sec:discuss}

The enormous, ionized wind in \ftm\ is a remarkable example of a quasar-driven outflow \citep{shen23}. It contrasts, however, with the spherical outflows surrounding ``normal'' Type 1 and 2 quasars at similar redshifts \citep{2013MNRAS.436.2576L,2013MNRAS.430.2327L,2014MNRAS.442.1303L} in its extreme velocities (reaching $W_{80\%} > 2000$~\kms) and dramatic, bipolar shape. Quasars with such high velocities are rare at $z \sim 0.5$ \citep{2014MNRAS.442..784Z, 2016MNRAS.461.3724W}, but more common at Cosmic Noon \citep{2016MNRAS.459.3144Z,2020A&A...634A.116V}. The \ftm\ wind bears a closer resemblance to 10-kpc scale bipolar outflows observed around radio-quiet, Type 1 and 2 quasars at lower redshifts \citep{2012ApJ...746...86G,2015ApJ...800...45H,2017ApJ...850...40R}.

\begin{figure*}
  \centering
  \includegraphics[width=\textwidth]{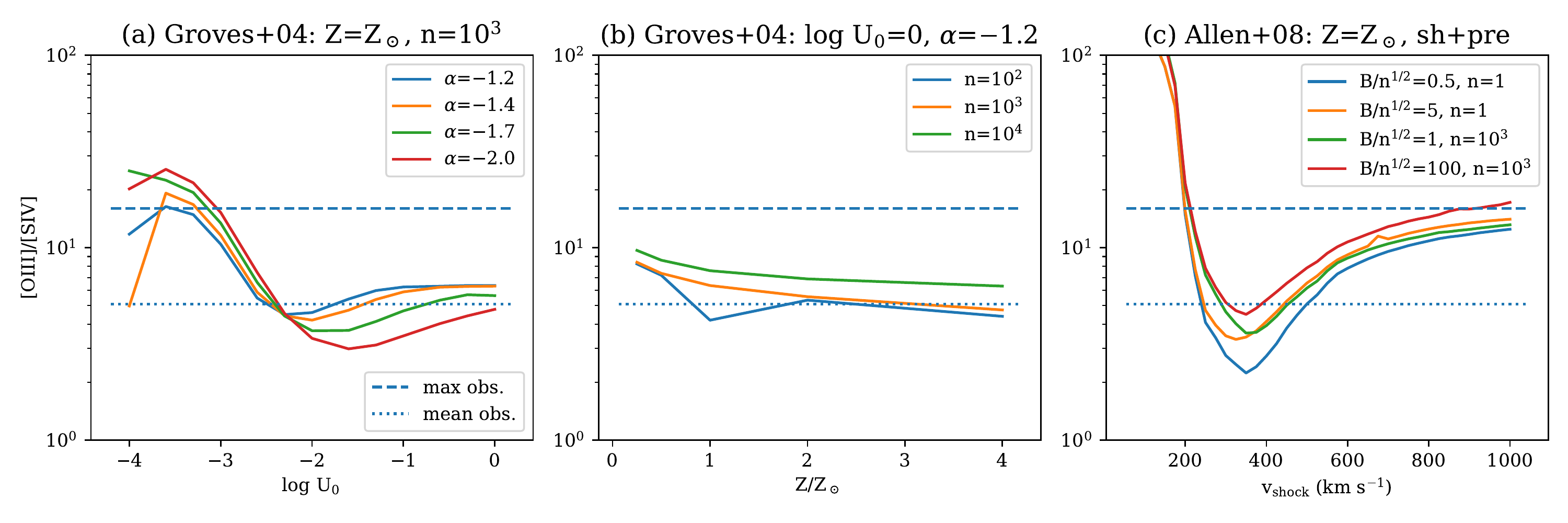}
    \caption{(a) \oth/\suf\ line ratio vs. ionization parameter $U_0$ and spectral index $\alpha$. These dusty, radiation-pressure-dominated photoionization models \citep{2004ApJS..153....9G} have $Z = Z_\odot$ and $n=10^3$~cm$^{-3}$. The dashed and dotted lines show the maximum and mean observed line ratios. The high-ionization models predict ratios 2--3$\times$ lower than the maximum observed line ratios. (b) \oth/\suf\ vs. metallicity and density for log $U_0 = 0$ and $\alpha=-1.2$. There is little variation with metallicity or density. (c) \oth/\suf\ vs. shock velocity for the shock$+$precursor models of \citet{2008ApJS..178...20A} with varying magnetic parameter $B/n^{1/2}$ (in units of $\mu$G~cm$^{-3/2}$) and density $n$ (in units of cm$^{-3}$). The lowest- and highest-velocity shock$+$precursor models are consistent with the maximum observed ratios. However, the low-velocity models cannot reproduce the observed higher-ionization Ne line fluxes.}
    \label{fig:groves-allen}
\end{figure*}

The close connection between \sufl\ and \othl\ is unsurprising given the similar conditions that should give rise to them, based on their ionization potentials and critical densities (Section~\ref{sec:intro}). AGN photoionization models bear out this expectation. Here we compare to the 1D equilibrium, plane-parallel models of \citet{2004ApJS..153....9G}, which explicitly incorporate the effects of dust mixed with the narrow-line region gas. The dust couples to the ionizing radiation, allowing for radial motion driven by radiation pressure. For dimensionless ionization parameters $U_0$ that would be expected in a quasar like \ftm\ (log~$U_0\sim-3$ to 0; \citealt{baskin05}), the line ratio \oth/\suf\ does not depend strongly on the shape of the ionizing continuum $\alpha$, on metallicity $Z$, or on density $n$ (Figure~\ref{fig:groves-allen}a--b). The ratio varies by a factor $\la3$ in the range of $Z/Z_\odot = 0.25-4.0$, $\alpha = -1.2$ to $-$2.0, and $n=10^2-10^4$~cm$^{-3}$. 

\begin{figure*}
  \centering
  \includegraphics[width=\textwidth]{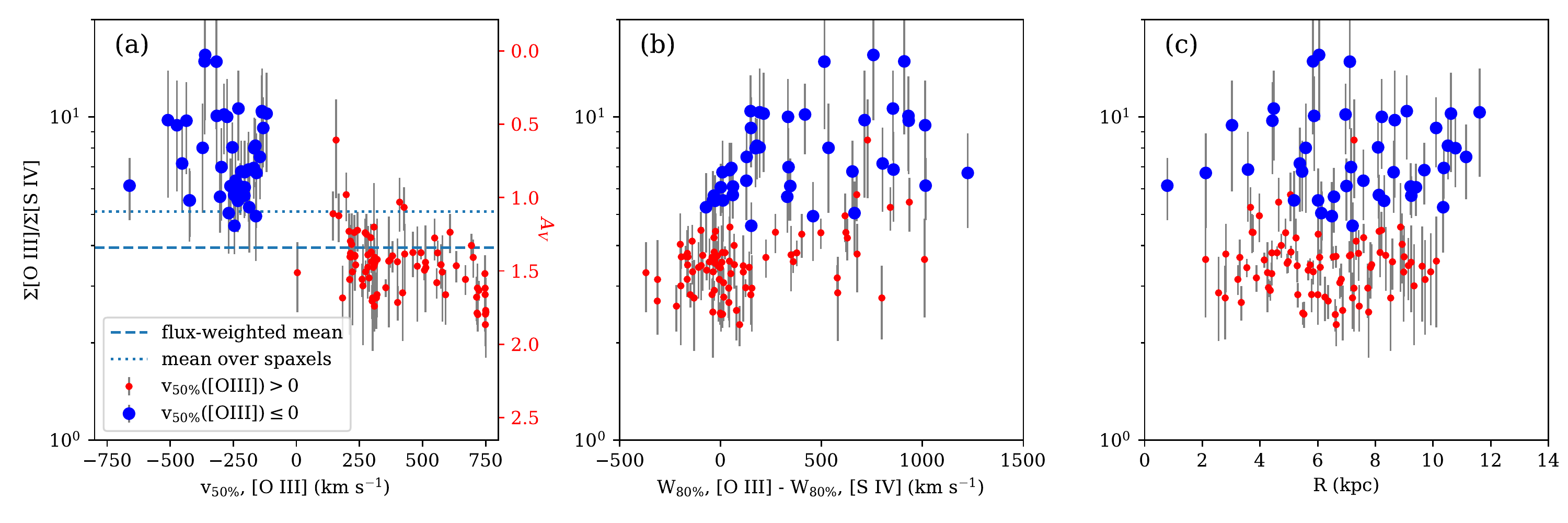}
    \caption{(a) \oth/\suf\ line ratio vs. \oth\ median velocity.  A significant correlation is evident ($r = -0.75$), suggesting differential attentuation with velocity. Spaxels with \oth\ $v_{50\%}>0$~\kms\ are in red; blueshifted spaxels are in blue. The right-side axis shows extrapolated extinction $A_V$, assuming an intrinsic line ratio of 16--the maximum observed in the data--and the \citet{2006ApJ...637..774C} extinction curve. Horizontal lines show average line ratios. (b) \oth/\suf\ vs. linewidth difference $W_{80\%}$(\oth) $-$ $W_{80\%}$(\suf); a milder correlation is present. (c) \oth/\suf\ vs. projected galactocentric radius. Although one might expect stronger differential extinction effects closer to the nucleus, no correlation is evident.}
    \label{fig:s4o3-vs}
\end{figure*}

Shock models show a wider range in predicted line ratios for these transitions. In Figure~\ref{fig:groves-allen}c, we compare solar metallicity models from \citet{2008ApJS..178...20A} with the observed ratios. The increase of the model line ratio with shock velocity above $v_\mathrm{shock} = 300$~\kms\ cannot drive the observed correlation with $v_{50\%}$, since then the ratio would depend on $|v_{50\%}|$; Figure~\ref{fig:s4o3-vs} would show a ``v'' shape rather than a straight line.

Further constraints arise from higher-ionization lines present in the 1D spectrum of the nebular peak. Table~\ref{tab:lines} lists various Ne flux ratios in this spectrum. The high-ionization lines ([\ion{Ne}{6}], [\ion{Ne}{5}]) exclude low-velocity shock solutions and also favor AGN photoionization over high-velocity shocks. They constrain photoionization models to ionization parameters log $U_0>-2$. The observed ratios are in line with those of Palomar-Green quasars and infrared-luminous quasars \citep{2009ApJS..182..628V}.

We also compare the Ne lines to \suf, which is the brightest observed line. Ne$^{+2}$ has an ionization energy of 40.96~eV, similar to S$^{+3}$. The \suf/[\ion{Ne}{3}] ratio is most consistent with AGN photoionization models with log $U_0>-2$. Both of these lines can be excited by stellar photoionization, but aside from compact dwarfs, the ratio is $\ll$1 \citep{2003A&A...403..829V}.

\begin{deluxetable}{cc}
    \tablecolumns{2}
    \tablecaption{Nebular Peak Line Ratios\label{tab:lines}}
    \tablewidth{0pt}
    \tablehead{
    \colhead{Quantity} & \colhead{Value} \\
    \colhead{(1)} & \colhead{(2)}
    }
  \startdata
  [\ion{Ne}{6}]7.65/[\ion{Ne}{2}]12.81 & 0.67$\pm$0.22 \\\relax
  [\ion{Ne}{6}]7.65/[\ion{Ne}{3}]15.55 & 0.55$\pm$0.17 \\\relax
  [\ion{Ne}{6}]7.65/[\ion{Ne}{5}]14.32 & 1.30$\pm$0.58 \\\relax
  [\ion{Ne}{5}]14.32/[\ion{Ne}{3}]15.55 & 0.51$\pm$0.19 \\\relax
  [\ion{Ne}{5}]14.32/[\ion{Ne}{2}]12.81 & 0.42$\pm$0.15 \\\relax
  [\ion{Ne}{3}]15.55/[\ion{Ne}{2}]12.81 & 1.20$\pm$0.25 \\\relax
  [\ion{S}{4}]10.51/[\ion{Ne}{3}]15.55 & 1.40$\pm$0.17 \\
    \hline
  [\ion{S}{4}]10.51 & (3.76$\pm$0.17)$\times$10$^{-16}$~erg~s$^{-1}$~cm$^{-2}$ \\
  \enddata

  \tablecomments{Line ratios and flux of the \sufl\ line in the 0\farcs6 aperture extracted at the nebular peak. Errors are 1$\sigma$.}
\end{deluxetable}

\begin{figure}
  \includegraphics[width=0.5\textwidth]{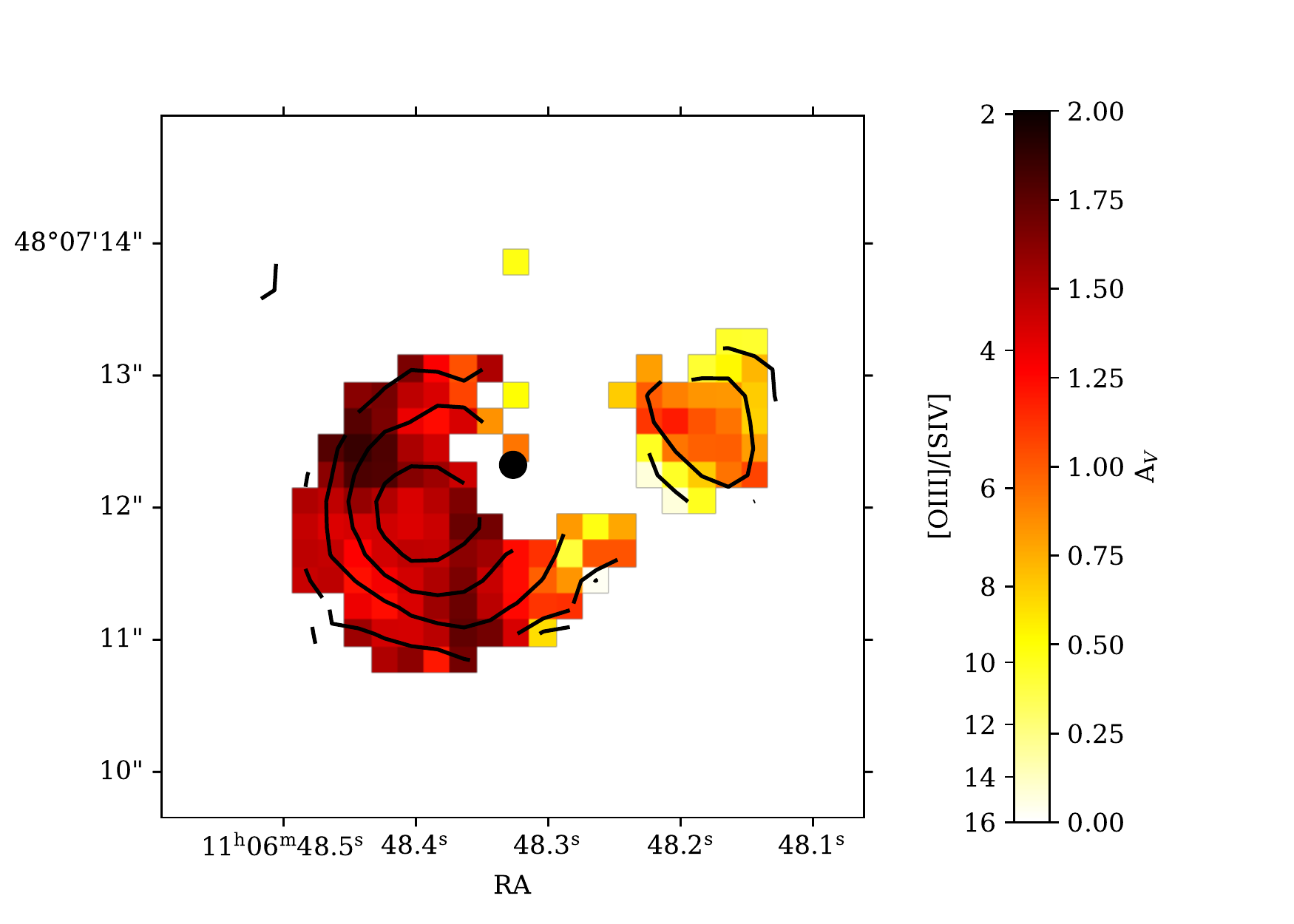}
    \caption{Map of \oth/\suf\ and $A_\mathrm{V}$, estimated from a model in which the intrinsic line ratio is near the maximum observed ($\sim$16) and smaller values are due to differential obscuration. The redshifted, receding side of the outflow is preferentially extincted due to obscuration by, e.g., the host galaxy. Extinctions are calculated using \citet{2006ApJ...637..774C}.}
    \label{fig:o3s4-av}
\end{figure}

Assuming a relatively constant \oth/\suf\ for the \ftm\ nebula, the factor-of-8 variation we observe (Figure~\ref{fig:s4-vs-o3}a) is thus likely caused by changing line-of-sight obscuration. This arises naturally in the bipolar outflow model, as the blueshifted side of the outflow is moving toward the observer on the nearer side of the system, and is thus less susceptible to dust attenuation than the redshifted side. Here we assume that the \suf\ and \oth\ line-emitting gases arise from the same physical locations and emit into the same solid angles. In this model, the intrinsic, unattenuated ratio is (\oth/\suf)$_\mathrm{int} \sim 16$, the highest observed in the nebula. Lower values can be mapped onto a screen visual extinction $A_V$ if we assume an extinction curve. Using \citet{2006ApJ...637..774C}, we find that the lowest observed ratios (\oth/\suf)$_\mathrm{int} = 2$ imply $A_V = 2.0$ mag. We show the distribution of line ratios and extinctions in Figure~\ref{fig:o3s4-av}.

In this scenario, the intrinsic, unattenuated line ratio---i.e., the maximum value observed in Figure~\ref{fig:groves-allen}---is 2--3$\times$ higher than predicted by dusty AGN photoionization models. These models best match the overall ionization conditions, as the presence of strong higher-ionization lines appears to rule out both slow (200~km~s$^{-1}$) and even very fast shocks (Figure~\ref{fig:groves-allen}). One possible culprit for this difference could be the uncertain sulfur depletion. Sulfur is assumed to be undepleted in the AGN models, while oxygen experiences a mild depletion of $-$0.22~dex \citep{2004ApJS..153....9G}. However, the actual depletion of sulfur in the diffuse interstellar medium (ISM) is uncertain \citep{2009ApJ...700.1299J,2009ApJ...693.1236C}, and it may be highly depleted in the cool, dense ISM and protoplanetary disks \citep{2019ApJ...885..114K}. A depletion of even $-$0.3~dex for sulfur could push the \oth/\suf\ ratio predicted by the photoionization models into agreement with our assumed, intrinsic value (unless the gas cooling is significantly affected by the change in sulfur abundance; \citealt{2023MNRAS.520.4345G}).

This scenario is entirely consistent with the integrated continuum extinction of $A_V = 1.2$ (Section~\ref{sec:intro}). The color excess $E(B-V)=0.44$ of \ftm\ is derived from a fit to the ratio of its optical-to-near-IR spectrum to that of a quasar composite that is attenuated by an Small Magellanic Cloud (SMC) extinction law \citep{2007ApJ...666..806N,2012ApJ...757...51G}. The visual extinction of the central quasar then results from applying an SMC-like differential extinction $R_V = 2.87$ \citep{1998ApJ...500..816G,2015MNRAS.453.3932R}

We observe a flux-weighted mean line ratio $(\oth/\suf) = 4$, which corresponds to $A_V=1.4$ mag, while the mean over spaxels is 5, or $A_V = 1.1$ mag (Figure~\ref{fig:s4o3-vs}a).  Examination of the preliminary full 3.5--19.5~$\mu$m quasar spectrum in our dataset does not show significant silicate absorption, which is consistent with other red and Type 2 quasars \citep{2012ApJ...757..125U,2008AJ....136.1607Z} and may be due to the geometry or clumpiness of the obscuring material.

Thus, the dust screen extincting the wind may also be responsible for extincting the quasar itself, as has been proposed for Type~2 and red, Type~1 quasars \citep{2014MNRAS.442..784Z, 2019MNRAS.488.3109K, 2021MNRAS.505.5283R,2022ApJ...934..119G}. Some observations suggest that the dusty winds extincting red quasars arises on scales $\la$10~pc \citep{2021A&A...649A.102C}, while others are consistent with kiloparsec scales \citep{2021MNRAS.505.5283R, 2021MNRAS.504.4445V}. If in fact dust in the large-scale outflow in \ftm\ also extincts the quasar, then the relevant scales are $\ga$1~kpc.

In this model, the tight correlation of \oth/\suf\ with \oth\ median velocity could reflect one of two possibilities. First, the flux ratio may imply that the extinction is not in fact strongly dependent on velocity. Rather, the correlation reflects the changing contribution of the red and blue sides of the outflow to the total flux. This in turn shifts the median velocity redward or blueward, depending on whether the more extincted red side dominates or not. Panels A and C in Figure~\ref{fig:s4-on-o3}d would tend to support this interpretation; they show two distinct velocity components that spatially overlap but have very different flux ratios on a component-by-component basis.

Second, there may be a dependence of extinction on velocity on each side of the bipolar flow. This view would be more consistent with the highest and lowest ratios being observed at the most extreme velocities. In fact, a correlation of the line ratio with velocity is seen within the parts of the nebula that do not show overlap between the two sides of the outflow (i.e., outside of the band of high \oth\ $W_{80\%}$, lending support to this idea. In reality, both of these effects may be at play. Finally, the lack of correlation with radius (Figure~\ref{fig:s4o3-vs}c) implies that the dust is not centrally concentrated, but rather spread throughout the extended, $r = 10$~kpc nebula or host disk by the outflow.

\section{Conclusions} \label{sec:conclude}

A key science goal of \jwst\ is assessing the impact of AGN feedback at Cosmic Noon, the era of peak star formation and AGN activity. Here we present the first mid-IR, integral-field data from the Early Release Science program Q3D, which aims to demonstrate the capabilities of \jwst\ to reach this goal. Our target, the red quasar \ftm, contains a fast, bipolar wind flowing to $\ga$10~kpc \citep{shen23}. This wind, as detected in \othl\ using ground-based, rest-frame optical integral-field spectroscopy, is an example of large-scale outflows in red quasars.

Mapping the morphology and kinematics of \oth\ using integral-field spectrographs has been one of the most transformative and productive techniques for understanding quasar-driven winds, their launching mechanisms and their reach into the quasar host and circumgalactic medium. Here we use \jwst\ MIRI observations to demonstrate for the first time that \sufl\ clearly traces the same phase of quasar outflows as \oth. 

With MIRI/MRS spectroscopy, we map the \suf\ emission line across the 20~kpc outflowing nebula in \ftm. We find that, at the peak of the outflow's emission, this is the most luminous line in the 3.5--19.5~$\mu$m spectral range. The S$^{+3}$ ion, and the \jwst\ observations that reveal it, are a close match to O$^{+2}$ and the \oth\ data in ionization potential (35~eV), sensitivity (10$^{-17}$~erg~s$^{-1}$~cm$^{-2}$~arcsec$^{-2}$), and spatial resolution (0\farcs5). The kinematics and morphology of the wind in F2M1106 seen in these two lines are remarkably similar.  The close alignment in line shape across much of the outflow means that the \suf\ line is an excellent proxy for \oth\ in the mid-IR.

The clear advantage of \suf\ is its lower sensitivity to extinction, particularly in the context of obscured AGN. However, \oth\ is intrinsically brighter. In red and obscured quasars, the optical continuum is strongly suppressed by the obscuration but whose power emerges as thermal radiation of AGN-heated dust grains at mid-IR wavelengths. Thus, in these galaxies, \oth\ is also less affected by the PSF, which is comparably brighter in the mid-IR. So \oth\ better probes the central regions where the two sides of the bipolar wind overlap.

Based on the \oth/\suf\ line ratio, and various higher-excitation Ne lines in a spectrum extracted at the peak of the nebula, the gas is excited by the quasar radiation field. The \oth/\suf\ ratio varies little in these models. Since the lines have very different wavelengths, their ratio then proves to be an excellent probe of extinction. We find a tight anticorrelation between \oth/\suf\ and central velocity $v_{50\%}$, which points to a model in which the redshifted, background side of the wind is more extincted than the blueshifted, foreground side. This dust could be in the nebula itself or in the host disk.

The extinction of the quasar itself ($A_V\sim1.2$) is consistent with being obscured by the dust in the wind. Dusty winds have been proposed as the reddening source for obscured quasars \citep{2014ApJ...787...65H, 2022ApJ...934..119G}. In this scenario, the current data imply the dust that is reddening \ftm\ is in parts embedded in the wind at kiloparsec scales.



\begin{acknowledgements}

This work is based in part on observations made with the NASA/ESA/CSA James Webb Space Telescope. The specific observations analyzed can be accessed via \dataset[https://doi.org/10.17909/xcfs-hx44]{doi:10.17909/xcfs-hx44}. The data were obtained from the Mikulski Archive for Space Telescopes at the Space Telescope Science Institute, which is operated by the Association of Universities for Research in Astronomy, Inc., under NASA contract NAS 5-03127 for JWST. These observations are associated with program \#DD-ERS-1335. Support for program \#DD-ERS-1335 was provided by NASA through a grant from the Space Telescope Science Institute, which is operated by the Association of Universities for Research in Astronomy, Inc., under NASA contract NAS 5-03127.

This work is also based in part on observations obtained at the international Gemini Observatory, a program of NSF’s NOIRLab, which is managed by the Association of Universities for Research in Astronomy (AURA) under a cooperative agreement with the National Science Foundation on behalf of the Gemini Observatory partnership: the National Science Foundation (United States), National Research Council (Canada), Agencia Nacional de Investigaci\'{o}n y Desarrollo (Chile), Ministerio de Ciencia, Tecnolog\'{i}a e Innovaci\'{o}n (Argentina), Minist\'{e}rio da Ci\^{e}ncia, Tecnologia, Inova\c{c}\~{o}es e Comunica\c{c}\~{o}es (Brazil), and Korea Astronomy and Space Science Institute (Republic of Korea). This work was enabled by observations made from the Gemini North telescope, located within the Maunakea Science Reserve and adjacent to the summit of Maunakea. We are grateful for the privilege of observing the Universe from a place that is unique in both its astronomical quality and its cultural significance.

\end{acknowledgements}

\software{\texttt{photutils} \citep{larry_bradley_2022_6825092}, \texttt{q3dfit} \citep{2014ascl.soft09005R,2017ApJ...850...40R}, \texttt{reproject} \citep{thomas_robitaille_2023_7950746}}
\facilities{Gemini:Gillett (GMOS-IFU), JWST (MIRI-MRS)}

\clearpage

\appendix
\section{Point-spread Function}
\label{sec:appendix}
\begin{figure}
    \centering
    \includegraphics[width=\textwidth]{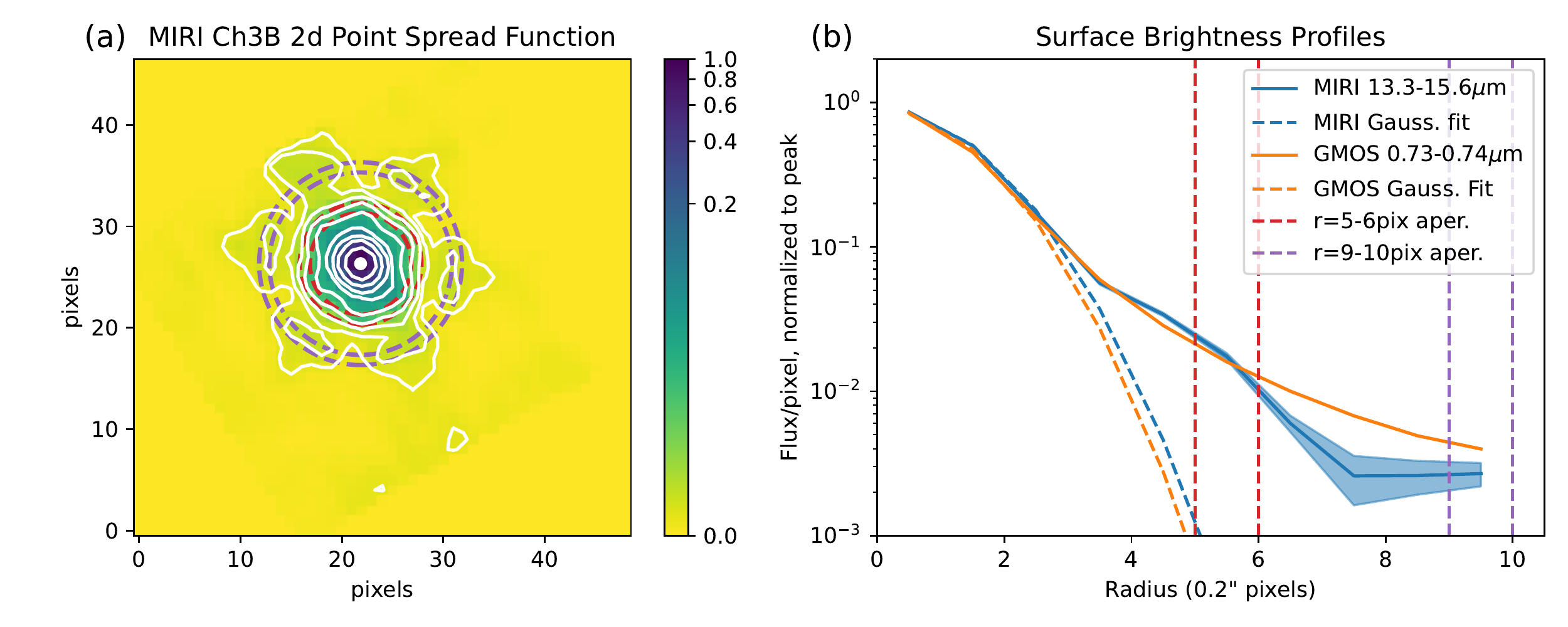}
    \caption{(a) MIRI Channel 3B point-spread function. White contours are spaced by factors of 2 in surface brightness from 0.512 times the peak flux to 0.002 of peak. The dashed red and purple lines mark two apertures used in the radial profile computation. (b) Radial surface brightness profiles of the MIRI and GMOS data, normalized to the 2D peak flux. The profiles are Gaussian within 3 pixels, with FWHM $=$ 0\farcs62-0\farcs65. They have non-Gaussian wings carrying $\sim$10-15\%\ power. The 1$\sigma$ error in the MIRI profile is represented by the shaded area. The apertures shown in (a) are delineated by the vertical dashed lines.}
    \label{fig:psf_comp}
\end{figure}

To assess the PSF of MIRI, we collapse the Channel 3B cube in wavelength (Figure~\ref{fig:psf_comp}a). The result shows the expected six-pointed azimuthal asymmetry. A radial surface brightness profile is an FWHM $=$ 0\farcs65 Gaussian within 3 pixels, with extended non-Gaussian wings that carry 10\%\ of the total power (Figure~\ref{fig:psf_comp}b; see also \citealt{2023A&A...675A.111A}.) The PSF asymmetry is most prominent only at very low flux levels far from the point source ($\la1\%$ of peak flux, at $\ga$1\arcsec). These wings are unimportant for the dynamic range we measure in the emission lines presented here.

For comparison, we show the GMOS radial profile (Figure~\ref{fig:psf_comp}b). The GMOS PSF is more azimuthally symmetric but almost identical to the MIRI PSF in radial shape.

\bibliography{f2m1106}
\bibliographystyle{aasjournal}

\end{document}